\documentstyle[12pt,epsfig]{article}
\begin{document}
\setlength{\parskip}{0.45cm}
\setlength{\baselineskip}{0.75cm}
\begin{titlepage}
\begin{flushright}
DO-TH 95/15 \\ RAL-TR-95-043  \\ August 1995
\end{flushright}
\vspace{0.5cm}
\begin{center}
\Large
{\bf Spin-dependent non-singlet structure functions} \\
\vspace{0.1cm}
{\bf in next-to-leading order} \\
\vspace{1.5cm}
\large
M. Stratmann, A. Weber* \\
\vspace{0.5cm}
\normalsize
Universit\"{a}t Dortmund, Institut f\"{u}r Physik, \\
\vspace{0.1cm}
D-44221 Dortmund, Germany \\
\vspace{1cm}
\large
W. Vogelsang \\
\vspace{0.5cm}
\normalsize
Rutherford Appleton Laboratory \\
\vspace{0.1cm}
Chilton Didcot, Oxon OX11 0QX, England \\
\vspace{2cm}
{\bf Abstract} \\
\end{center}
We study in detail the flavor-non-singlet component of polarized
structure functions in the framework of a consistent and complete
next-to-leading order (${\cal O}(\alpha_s))$ analysis. In this context,
we discuss some important features of the calculation of the
next-to-leading order corrections. Particular emphasis is put on the
$Q^2$-evolution of sum-rules for the first moments of the non-singlet
structure functions which, as we show, could serve to explore $SU(2)$ and
$SU(3)$ breaking effects in relations between baryonic $\beta$-decay matrix
elements and in the proton's polarized sea. Furthermore we make
predictions for polarized non-singlet structure functions
measurable in a conceivable $\vec{e}\vec{p}$ collider mode of HERA.
\begin{description}
\item[*] Now at: Koninklijke/Shell Exploratie en Produktie
Laboratorium, P.O. Box 60,\linebreak
\hspace*{1.22cm} NL-2280 AB Rijswijk, The Netherlands
\end{description}
\end{titlepage}
\section{Introduction}
The spin-dependent structure functions of protons, neutrons, and
deuterons have received much attention both experimentally and
theoretically in the past years. Since the advent of the EMC
result \cite{emc} on the proton's $g_1^p(x, \langle Q^2 \rangle
=10.7 \mbox{\rm GeV}^2)$, most theoretical studies have been
focussed on the singlet component of this structure function
in order to explain its unexpected experimental smallness,
hereby assuming that the non-singlet (NS) component is rather
well understood. First experimental evidence for this latter
assumption was provided recently by the confirmation \cite{smc,e143}
of the Bj\o rken sum-rule \cite{bj} which relates the integrals
(first moments) of $g_1^p$ and $g_1^n$. However, this sum-rule,
which depends merely on the fundamental $SU(2)$ isospin
($u \leftrightarrow d$) symmetry
between matrix elements of charged and neutral axial currents and
is therefore expected to hold, does not entirely fix the
first moment of the NS
component, $g_{1,NS}^p$, of $g_1^p$, since the latter can be
written (in leading order (LO)) as
\begin{equation}
g_{1,NS}^p = \frac{1}{12}\Delta A_3 (1) + \frac{1}{36} \Delta A_8 (1) \:\:\:,
\end{equation}
where in terms of the first moments of the polarized (anti)quark densities
$\Delta \!\!\stackrel{\scriptscriptstyle (-)}{q}\!(x,Q^2)$ we have
\begin{eqnarray}
\Delta A_3 (1) &=&\int_0^1 \left( \Delta u+\Delta \bar{u}-
\Delta d-\Delta\bar{d} \right) (x,Q^2) dx  \:\:\:  , \nonumber \\
\Delta A_8 (1) &=&\int_0^1 \left( \Delta u+\Delta \bar{u}+\Delta d+
\Delta\bar{d} -2 (\Delta s+\Delta \bar{s}) \right) (x,Q^2) dx \:\:\: .
\end{eqnarray}
The Bj\o rken sum-rule \cite{bj} is equivalent to
\begin{equation}
\Delta A_3 (1) =F+D = g_A = 1.2573 \pm 0.0028 \:\:\: ,
\end{equation}
but information on $\Delta A_8$ can only be obtained from hyperon
$\beta$-decays. Assuming full $SU(3)$ symmetry between hyperon
decay matrix elements of the flavor-changing weak axial currents and
the neutral ones, one finds (with $F$, $D$ taken from ref.\ \cite{close})
\begin{equation}
\Delta A_8 (1) =3F-D=0.579 \pm 0.025 \:\:\: .
\end{equation}
This approach has been seriously
questioned in ref.\ \cite{lip}, where the suggestion
was made that $SU(3)_f$ symmetry is broken in such a way that
only the {\em valence} $\Delta q_v=\Delta q - \Delta \bar{q}$ content of
$\Delta A_3 (1)$, $\Delta A_8 (1)$,
rather than the full combinations $\Delta A_3 (1)$, $\Delta A_8 (1)$,
enters eqs.\ (3),(4). In view of these uncertainties and of the fact that
the baryonic $\beta$-decays cannot tell us anything about $g_{1,NS}^p$
except for the first moment, it is interesting to examine the
NS sector of polarized structure functions in more detail in order to
find other possible experimental clues to $\Delta A_3$, $\Delta A_8$,
and the polarized valence densities $\Delta u_v$, $\Delta d_v$,
hereby improving our present understanding of the relation between the first
moments of these quantities and the $F$ and $D$ values. In this respect,
it is necessary to consider not only the NS piece of the
electromagnetic (e.m.)
structure functions $g_1^{p,n}\equiv g_1^{ep,en}$, but also the
polarized electroweak structure functions $g_3$, $g_4$ studied in
refs.\ [7-14], which partly are pure NS quantities. Since possible
measurements of such structure functions are likely to be performed at
$Q^2$ much higher than those relevant for eqs.\ (3),(4), it is important
to theoretically understand the $Q^2$-evolution of
spin-dependent NS structure
functions as well as possible. For this purpose, it is necessary to improve
the theoretical predictions in the NS sector, by performing a complete and
consistent next-to-leading order (NLO) analysis. All ingredients
for this are available as we will see below, and there are some features
of the NLO corrections which are interesting in themselves. Also, the
theoretical predictions are much more reliable for the NS sector since
it is not plagued by the anomaly contribution as is the case for the
singlet contributions to polarized structure functions \cite{ar}.

The remainder of this paper is organized as follows: In Section 2
we review the main LO results on spin-dependent NS structure
functions. Section 3 gives a detailed account of the determination of the
NLO corrections. Section 4 contains the numerical evaluation of
our results, and in Section 5 we summarize our findings.
\section{Spin-dependent non-singlet structure functions in leading order}
The structure functions $g_1$, $g_3$, and $g_4$ appear in the hadronic
tensor as (see, e.g., \cite{wir})
\begin{equation}
W_{\mu\nu} = - i\epsilon_{\mu\nu\rho\sigma}
\frac{q^{\rho}p^{\sigma}}{p \cdot q}g_1+\left( -g_{\mu\nu}+
\frac{q_{\mu}q_{\nu}}{q^{2}}\right) g_3
+ \frac{1}{p\cdot q}\left( p_{\mu}-\frac{p\cdot q}{q^{2}}q_{\mu}\right)
\left( p_{\nu}-\frac{p\cdot q}{q^{2}}q_{\nu} \right) g_4  \:\:\: ,
\end{equation}
where we have already replaced $s^{\mu}\rightarrow p^{\mu}$ for the
spin vector of a longitudinally polarized nucleon with momentum $p$.
In eq.\ (5), $q$ denotes the momentum of the virtual boson probing the
hadron. As is well-known, $g_3$ and $g_4$ do not contribute to purely
e.m. scattering, but appear in (parity-violating) electroweak
neutral current (NC) or charged current (CC) lepton-nucleon interactions.
Therefore, their experimental accessibility may seem remote presently.
However, they could be measured in
$\stackrel{\scriptscriptstyle (-)}{\nu}$-scattering
off a polarized target, and they would certainly play a role in deep-inelastic
scattering (DIS) experiments at HERA if also the 820 GeV proton beam could be
longitudinally polarized \cite{barber}.
All relevant cross section formulas for
$\stackrel{\scriptscriptstyle (-)}{\nu}\!\! \vec{p}$,
$e^{\pm}\vec{p}$ interactions in terms of $g_1$, $g_3$, and $g_4$
can be found, e.g., in refs.\ \cite{wir,ans} and need not be repeated here. As
was shown in \cite{wir}, the LO expressions for the structure functions
can be cast into the forms
\begin{eqnarray}
g_{1}(n,Q^{2})&=&\frac{1}{2}\sum_{q} S_{q} \left( \Delta q(n,Q^{2})+\Delta
\bar{q}(n,Q^{2}) \right) \nonumber \\
g_{3}(n,Q^{2})&=&\frac{1}{2}\sum_{q} R_{q} \left( \Delta q(n,Q^{2})-\Delta
\bar{q}(n,Q^{2}) \right) \\
g_{4}(n-1,Q^{2})&=&\sum_{q} R_{q} \left( \Delta q(n,Q^{2})-\Delta \bar{q}
(n,Q^{2}) \right) =2 g_3 (n,Q^{2} ) \:\:\: , \nonumber
\end{eqnarray}
where, as usual, the Mellin-$n$ moments of a Bj\o rken-$x$-dependent
function $g(x)$ are defined as $g(n)=\int_0^1 x^{n-1} g(x) dx$.
The coefficients
$S_q$ and $R_q$ in eq.\ (6) depend on the exchanged boson, $\gamma^{\ast}$,
$Z^0$, $W^{\pm}$, in the DIS process and can be found in \cite{wir}.
Obviously, for $W^{\pm}$-exchange (CC interactions), only the quark {\em or}
the antiquark of a given flavor contributes, depending on the charge
of the $W$. To become more specific, we write the various conceivable
structure functions in LO in terms of the NS quark
combinations (for eqs.\ (7-20)
below we drop the obvious argument $(n,Q^2)$ from all quantities)
\begin{eqnarray}
&&\Delta u_v = \Delta u -\Delta \bar{u} \:\: , \:\:\:\:\:
\Delta d_v = \Delta d -\Delta \bar{d} \:\:\: , \nonumber \\
&&\Delta A_3 = \Delta u +\Delta \bar{u} - \Delta d - \Delta \bar{d}
\:\:\: , \nonumber \\
&&\Delta A_8 =\Delta u +\Delta \bar{u} + \Delta d +
\Delta \bar{d}-2 (\Delta s+\Delta \bar{s} )
\end{eqnarray}
and the singlet
\begin{equation}
\Delta \Sigma =\Delta u +\Delta \bar{u} + \Delta d +
\Delta \bar{d} + \Delta s+\Delta \bar{s}
\end{equation}
(for $f=3$ flavors) as \cite{wray}
\begin{eqnarray}
g_1^{ep \: (e.m.)} &=& \frac{1}{12}\Delta A_3  +
\frac{1}{36} \Delta A_8  +\frac{1}{9} \Delta \Sigma    \\
g_3^{ep,NC} &=& \frac{1}{4} \left(
\frac{2}{3} \Delta u_v  + \frac{1}{3} \Delta d_v \right)
\end{eqnarray}
for $\vec{e}\vec{p}$ NC interactions, where for $g_1^{ep}$ we have only
written the purely electromagnetic contribution (which dominates \cite{wir}
if polarized electrons are used) since the other NC contributions do not
easily lead to NS quantities. $g_3^{ep,NC}$ has been written only for
the dominant \cite{wir} contribution from $\gamma Z^0$
interference\footnote{
As compared to ref.\ \cite{wir} we have dropped a factor
$\left( 4\sin^2 \Theta_W \cos^2 \Theta_W (Q^2+M_Z^2)/Q^2 \right)^{-1}$
from the normalization of $g_3^{ep,NC}$, where $\Theta_W$ is the Weinberg
angle and $M_Z$ the $Z^0$ mass.}.
For CC structure functions
($\stackrel{\scriptscriptstyle (-)}{\nu}\!\! p\rightarrow
e^{\pm}X$ or $e^{\pm}p\rightarrow
\stackrel{\scriptscriptstyle (-)}{\nu}\!\! X$ scattering)
one has \cite{wir}
\begin{eqnarray}
g_{1}^{\nu p}\!\! &=& \!\! \Delta d  + \Delta \bar{u}+
f_1 (\lambda) \Delta s = \frac{1}{2} \left( \Delta d_v -\Delta u_v
+\Delta \Sigma  \right) -\frac{1}{6}(1-f_1(\lambda)) \left(
\Delta \Sigma  -\Delta A_8 \right) \\
g_{1}^{\bar{\nu} p}\!\! &=& \!\!\Delta u + \Delta \bar{d} +
f_1(\lambda) \Delta \bar{s}
=\frac{1}{2} \left( \Delta u_v -\Delta d_v +\Delta \Sigma  \right)
-\frac{1}{6}(1-f_1(\lambda)) \left( \Delta \Sigma  -\Delta A_8 \right) \\
g_{3}^{\nu p}\!\! &=& \!\!-\Delta d  + \Delta \bar{u}-
f_3(\lambda)\Delta s
= \frac{1}{2} \left( \Delta A_3  - \Delta u_v
-\Delta d_v \right) -\frac{f_3(\lambda)}{6} \left( \Delta \Sigma
-\Delta A_8  \right) \\
g_{3}^{\bar{\nu} p} \!\! &=& \!\! -\Delta u + \Delta \bar{d} +
f_3(\lambda) \Delta \bar{s} =
-\frac{1}{2} \left( \Delta A_3  + \Delta u_v
+\Delta d_v \right) +\frac{f_3(\lambda)}{6} \left( \Delta \Sigma
-\Delta A_8  \right) \:\:,
\end{eqnarray}
where we have introduced functions $f_i (\lambda=Q^2/(Q^2+m_c^2))$
with $f_i (1)=1$ (massless limit) which take fully into account the
effects of the charm
mass $m_c$ in the $s\rightarrow c$ transition. In the LO considered
in eqs.\ (11-14), the $f_i (\lambda)$ are simply given by the 'slow-rescaling'
prescription \cite{barn} which yields $x_{Bj} = Q^2/2pq
\rightarrow x_{Bj}/\lambda$ and therefore $f_1 (\lambda)=
f_3 (\lambda) \equiv f(\lambda) = \lambda^n$ for the
$n$th moment in eqs.\ (11-14). The expressions for the $n$th moment
of the structure function $g_4/2x$, $g_4 (n-1,Q^2)/2$, are the same
\cite{wray,derm,lampe,wir} as the
right-hand-sides (rhs) of eqs.\ (10),(13),(14) with, however \cite{wir},
$f_4 (\lambda) = f (\lambda)/\lambda$ in the CC case. In this way
one finds a (slight) violation of the Callan-Gross-like relation
$g_4(n-1,Q^2)=2 g_3(n,Q^2)$ by terms of ${\cal O}(m_c^2/Q^2)$ due to
the CC $s\rightarrow c$ transitions. Note that eqs.\ (11-14) can be
easily seen to receive only corrections of ${\cal O}(\sin^2 \theta_c
m_c^2/Q^2)$ when taking into account the effects of Cabibbo-mixing;
we can safely neglect these small terms.
The structure functions for DIS scattering off neutron-targets can
be easily obtained by changing the signs of the $\Delta A_3$ terms
and interchanging
$\Delta u_v \leftrightarrow \Delta d_v$ in eqs.\ (9-14). From
eqs.\ (9-14) we can, e.g., construct the following NS combinations
\cite{wray}:
\begin{eqnarray}
g_1^{ep\: (e.m.)} - g_1^{en \: (e.m.)}  &=&
\frac{1}{6} \Delta A_3 \\
g_1^{\bar{\nu}p} - g_1^{\nu p}  &=&
\Delta u_v  -\Delta d_v  \\
g_3^{\nu p} - g_3^{\nu n} &=& \Delta A_3  \\
g_3^{\bar{\nu}p} + g_3^{\nu p} &=& -\left( \Delta u_v + \Delta d_v
\right) \\
9 \left( g_1^{ep \: (e.m.)} + g_1^{en (e.m.)}  \right) &-&
\frac{6}{2+f (\lambda)} \left( g_1^{\nu p} + g_1^{\nu n}  \right)
= \frac{1}{2} \frac{5f (\lambda)-2}{f (\lambda)+2} \Delta A_8  \\
g_1^{\bar{\nu}p} + g_1^{\nu p}  - \frac{2+f (\lambda)}{f (\lambda)}
\left( g_3^{\bar{\nu} p} - g_3^{\nu n} \right)
&=& \Delta A_8\;\;\;,
\end{eqnarray}
etc. Besides these relations, the NC $g_3^{ep,NC} (n,Q^2)$ in eq.\
(10) is obviously also an entire NS quantity.
\section{Next-to-leading order corrections}
In order to study the evolution of the polarized NS structure functions in
NLO it is necessary to recall the well-known solution (see, e.g., \cite{gr})
of the NS renormalization group equation relating the Mellin-$n$ moments
of a polarized NS structure function $\tilde{g}_{i,NS}=g_1$, $g_3$,
$g_4/2x$ at the input scale $Q_0^2$ and at $Q^2>Q_0^2$:
\begin{eqnarray}
\tilde{g}_{i,NS}(n,Q^2) &=&
\left( 1+\frac{\alpha_s (Q^2)-\alpha_s (Q_0^2)}{4\pi}
\left[ 2 \Delta C_q^i (n) + \frac{\Delta \gamma_{NS}^1 (n)}{2\beta_0}
-\frac{\beta_1 \gamma_{NS}^0 (n)}{2\beta_0^2} \right] \right)  \nonumber \\
&& \times
\left( \frac{\alpha_s (Q^2)}{\alpha_s (Q_0^2)}
\right)^{\gamma_{NS}^0(n)/2\beta_0} \tilde{g}_{i,NS}(n,Q_0^2)  \\
&=& \left( 1+\frac{\alpha_s (Q^2)}{2\pi} \Delta C_q^i (n) \right)
\Delta q_{NS}^i (n,Q^2)  \:\:\: ,
\end{eqnarray}
where
\begin{equation}
\frac{\alpha_s(Q^2)}{4\pi} \simeq \frac{1}{\beta_0
\ln Q^2/\Lambda_{\overline{\rm{MS}}}^2}-
\frac{\beta_1}{\beta_0^3} \frac{\ln\ln Q^2/\Lambda_{\overline{\rm{MS}}}^2}
{\left( \ln Q^2/\Lambda_{\overline{\rm{MS}}}^2\right)^2}
\end{equation}
with the QCD scale parameter
$\Lambda_{\overline{\rm{MS}}}$,
$\beta_0 = 11-2f/3$, $\beta_1=102-38f/3$, and
\begin{eqnarray}
\nonumber
\Delta q_{NS}^i (n,Q^2) &=& \left( 1+\frac{\alpha_s (Q^2)-
\alpha_s (Q_0^2)}{4\pi}\!
\left[ \frac{\Delta \gamma_{NS}^1 (n)}{2\beta_0}
-\frac{\beta_1 \gamma_{NS}^0 (n)}{2\beta_0^2} \right] \right)\\
&&\times
\left( \frac{\alpha_s (Q^2)}{\alpha_s (Q_0^2)}\right)^{
\gamma_{NS}^0(n)/2\beta_0}\! \Delta q_{NS}^i (n,Q_0^2)
\end{eqnarray}
is the suitable NS combination of polarized quark densities
evolved from $Q_0^2$
to $Q^2$ via the LO (one-loop) and NLO (two-loop) NS anomalous
dimensions\footnote{For the
perturbative expansions of the QCD-$\beta$-function
and the anomalous dimensions we use the expansion
parameter $\alpha_s/4\pi$ (see, e.g., \cite{gr} for detailed
expressions). Note also that we have omitted the '$\Delta$' for
the (polarized) $\gamma_{NS}^0(n)$ since this quantity is trivially equal
to its unpolarized counterpart \cite{ahr,ap}.}
$\gamma_{NS}^0 (n)$ and $\Delta \gamma_{NS}^1 (n)$. The
precise form of the NLO pieces ($\Delta C_q^i$, $\Delta \gamma_{NS}^1$) in
eqs.\ (21), (22) and (24) depends
on the factorization scheme convention adopted for the relation (22) between
the NS structure function and the relevant NS quark densities beyond the
leading order. Since the $\tilde{g}_{i,NS}$ are physical, i.e.\ measurable,
quantities and the $\gamma_{NS}^0 (n)$ \cite{ahr,ap} is
convention-independent, it becomes evident from eq.\ (21) that the scheme
dependences of $\Delta C_q^i (n)$
and $\Delta \gamma_{NS}^1(n)$ cancel each other
such that the combination $2 \Delta C_q^i (n)+\Delta \gamma_{NS}^1 (n)/2
\beta_0$ is scheme {\em independent} \cite{gr}.
Needless to say that removing all NLO quantities ($\Delta C_q^i$,
$\Delta \gamma_{NS}^1$, $\beta_1$) in eqs.\ (21-24),
we recover the LO results of eqs.\ (9-14) with the quark
density combinations evolving according to the LO
($\gamma_{NS}^0$ \cite{ahr,ap}) NS evolution equation.

Both the essential ingredients for the NLO calculation,
$\Delta C_q^i (n)$ and $\Delta \gamma_{NS}^1 (n)$, can be found in the
literature. To facilitate the further discussion, let us first turn
to the first moments, $n=1$, which are of particular interest in the
polarized case. As was discussed in ref.\ \cite{kod} in the framework
of the OPE, the operator corresponding to the first moments
$\Delta A_3 (1,Q^2)$ and $\Delta A_8 (1,Q^2)$ is nothing but the NS axial
current which is a conserved quantity and thus has vanishing anomalous
dimensions to all orders, which in particular means $\Delta \gamma_{NS}^1
(1)=0$. Furthermore, the value of the first moment
of the Wilson coefficient $\Delta C_q^1 (n)$ for $g_1^{ep \: (e.m.)}$
was found in \cite{kod,kod2} to be $\Delta C_q^1 (1)=-3 C_F/2$, giving
$-3 C_F$ for the scheme independent combination
$2 \Delta C_q^i (1)+\Delta \gamma_{NS}^1 (1)/2 \beta_0$ and,
according to eq.\ (22), leading to the factor $(1-\alpha_s/\pi)$ in the NS
sector of $g_1$. Of course,
both $\Delta C_q^1 (n)$ and $\Delta \gamma_{NS}^1 (n)$ depend on
the regularization scheme adopted in their calculation\footnote{
Needless to say that this has to be the same in the calculations
of these quantities.}, and different schemes will in principle
give different answers even for the first moments $\Delta
C_q^1 (1)$, $\Delta \gamma_{NS}^1 (1)$, though still respecting the
condition $2 \Delta C_q^1 (1)+\Delta \gamma_{NS}^1 (1)/2 \beta_0=
-3 C_F$. However, the conservation of the NS axial current {\em dictates}
the vanishing of $\Delta \gamma_{NS}^1 (1)$, and hence the value
$\Delta C_q^1 (1)=-3 C_F/2$, which means that a scheme transformation
has to be performed if these results are not automatically respected
by the regularization scheme used.

Let us briefly list the results obtained for $\Delta C_q^1 (1)$
(to be calculated in the process $\vec{\gamma}^{\ast} \vec{q}\rightarrow
q (g)$ to ${\cal O} (\alpha_s)$) using various regulators also
used previously in the corresponding calculations in the unpolarized
case. The result of ref.\ \cite{kod},
$\Delta C_q^1 (1)=-3 C_F/2$, was found using massless but
off-shell incoming quarks and on-shell outgoing gluons.
The same result for $\Delta C_q^1 (1)$ is obtained \cite{guil} if
one uses massless on-shell quarks, but off-shell gluons ($k^2 >0$).
Turning to dimensional regularization in $\overline{\mbox{\rm MS}}$,
the result depends on how the Dirac matrix $\gamma_5$, appearing
due to the projector on the quark's helicity, is treated in $n \neq 4$
dimensions. The prescription of a totally anticommuting $\gamma_5$ by
Chanowitz et al.\ \cite{cfh} yields
\cite{rat,alex} again\footnote{The same result in dimensional regularization
was found earlier in \cite{kod2} without specifying the
$\gamma_5$-prescription.}
$\Delta C_q^1 (1)=-3 C_F/2$. The same result is obtained
\cite{willi} in the closely related $\gamma_5$ scheme of ref.\ \cite{korn},
taking the $\gamma^{\ast}q$ vertex as the 'reading point' to be defined
in that scheme. However, when using the original scheme of 't Hooft and
Veltman \cite{tv} and Breitenlohner and Maison \cite{bm} (HVBM),
or the equivalent prescription of refs.\ \cite{del,lv}, one obtains
\cite{alex,phd}
\begin{equation}
\Delta C_q^1 (1)=-\frac{7}{2} C_F  \:\:\: ,
\end{equation}
(naively) corresponding to a correction $(1-7\alpha_s/3\pi)$ in the
NS sector of $g_1$ and to a non-zero value for the anomalous
dimension, $\Delta \gamma_{NS}^1 (1)$, in contradiction to the conservation
of the NS axial vector current. Finally, the same result,
$\Delta C_q^1 (1)=-7 C_F/2$, is obtained for massive on-shell
quarks ($m_q \neq 0$) in the process $\vec{\gamma}^{\ast}\vec{q}
\rightarrow q (g)$. For completeness, we list all the results
for $\Delta C_q^1 (n)$ for arbitrary Mellin-$n$ in the Appendix.
Comparing with the corresponding results [33-37]
for the Wilson coefficient
$C_q^2 (n)$ for the unpolarized structure function $F_2/2x$ in the
various regularizations, one finds that {\em all} the $\Delta C_q^1 (n)$
with the property $\Delta C_q^1 (1)=-3 C_F/2$ satisfy
\begin{equation}
C_q^2 (n) - \Delta C_q^1 (n) =  C_F \left( \frac{1}{n} + \frac{1}{n+1}
\right)
\end{equation}
(corresponding to $C_q^2 (z)-\Delta C_q^1 (z)=C_F (1+z)$ in Bj\o rken-$x$
space). This implied regularization scheme independence of
$C_q^2 (n) - \Delta C_q^1 (n)$ can be understood as follows: As
a consequence of the factorization theorem, the difference
$C_q^{DY} (n)-2 C_q^2 (n)$, where $C_q^{DY} (n)$ are the $n$-moments of the
${\cal O}(\alpha_s)$ (NS) quark corrections to the unpolarized
Drell-Yan process $q\bar{q}\rightarrow \gamma^{\ast}(g)$,
has to be the same in {\em any} scheme (see, for example, \cite{aem1,aem2}).
The same is true (see \cite{rat,alex}) for the difference
$\Delta C_q^{DY} (n) - 2 \Delta C_q^1 (n)$ with the
${\cal O}(\alpha_s)$ quark corrections $-\Delta C_q^{DY} (n)$
to the {\em polarized} Drell-Yan process $\vec{q}\vec{\bar{q}}
\rightarrow \gamma^{\ast}(g)$. On the other hand, the
annihilating quark lines in this process trivially give $\Delta C_q^{DY}(n)=
C_q^{DY}(n)$ {\em if} the regularization scheme used in the calculation
of $\Delta C_q^{DY}(n)$ respects chirality conservation.
It then automatically follows that $C_q^2 (n) - \Delta C_q^1 (n)$
is also the same in all such schemes\footnote{Alternatively, one can see
the expected scheme invariance of $C_q^2 (n) - \Delta C_q^1 (n)$ from
the fact that \cite{aem1,aem2}
$C_q^2 (n) - C_q^3 (n)= C_F \left( \frac{1}{n}+\frac{1}{n+1} \right)$
is scheme invariant (where $C_q^3(n)$ is the coefficient function of the
unpolarized structure function $F_3$) and from the similar appearance
of $F_3$ and $g_1$ in the hadronic tensor.}.

In contrast to eq.\ (26), we have for the calculation in the HVBM scheme
and for the $m_q \neq 0$ calculation (which gave
$\Delta C_q^1 (1)=-7 C_F/2$)
\begin{equation}
C_q^2 (n) - \Delta C_q^1 (n) =  C_F \left( \frac{1}{n} + \frac{1}{n+1}
\right) + \left\{  \begin{array}{ll}
                  4 C_F \left( \frac{1}{n} - \frac{1}{n+1} \right)
                  & \mbox{HVBM}  \\
                  2 C_F \frac{1}{n}  & (m_q \neq 0) \:\:\: .
                 \end{array}
\right.
\end{equation}
According to our previous observations, these regulators then necessarily
break the relation $\Delta C_q^{DY}(n)=C_q^{DY}(n)$, i.e., break
chirality. In fact, it was shown in ref.\ \cite{alex} that
\begin{equation}
\Delta C_{q,HVBM}^{DY} (n) = C_q^{DY} (n) - 8 C_F \left(
\frac{1}{n} - \frac{1}{n+1} \right)
\end{equation}
in the HVBM scheme, showing how the terms $\sim (1/n - 1/(n+1))$
in eqs.\ (27),(28) cancel out in the difference
$\Delta C_q^{DY} (n) - 2 \Delta C_q^1 (n)$, but individually break
the relations $\Delta C_q^{DY}(n)=C_q^{DY}(n)$ and
$C_q^2 (n) - \Delta C_q^1 (n) = C_F ( 1/n  + 1/(n+1) )$. Furthermore,
the terms $\sim (1/n - 1/(n+1))$ in the HVBM scheme originate
\cite{alex} from a configuration where the incoming and the outgoing
particles in the process $\gamma^{\ast}q\rightarrow qg$ become
collinear, and thus should rather be understood as part of the polarized
(NLO) quark densities. Finally, as far as the massive calculation
is concerned, the term $2 C_F/n$ after the curly bracket in eq.\ (27)
can be traced back to have its origin in a chirality breaking
term $\sim m_q^2$ which survives the eventual limit $m_q \rightarrow 0$
since it happens to be multiplied by a double-pole term.
Having found the origin of the additional terms in eq.\ (27) which
lead to $\Delta C_q^1 (1)=-7 C_F/2$, we expect that similar terms
would be present in the $\Delta \gamma_{NS}^1 (n)$ when calculated in the
HVBM or the massive scheme, such that scheme transformations,
by means of $2 \Delta C_q^1 (n)+\Delta \gamma_{NS}^1 (n)/2 \beta_0
= {\rm{invariant}}$, could be performed to eliminate these terms
from both $\Delta C_q^1$ and $\Delta \gamma_{NS}^1 (n)$.
Hereby one would obtain the correct values $\Delta C_q^1 (1)=-3 C_F/2$
and $\Delta \gamma_{NS}^1 (1)=0$ (as dictated \cite{kod} by the the
conservation of the NS axial current), and the relation
$C_q^2 (n) - \Delta C_q^1 (n) =  C_F ( 1/n  + 1/(n+1) )$ would be restored
in each case.

To complete the discussion of the Wilson coefficients $\Delta C_q^i$, let us
specify our final choices for the coefficients for $g_1$,
$g_3$, and $g_4$. Since, as we will see below, the anomalous dimension
$\Delta \gamma_{NS}^1 (n)$ is known within dimensional regularization
in the $\overline{\mbox{MS}}$-scheme, we have to choose the Wilson
coefficients accordingly. This means that the coefficient in (A.4)
\cite{kod2,rat,willi}
(or the one in (A.5) after elimination of the chirality breaking term
$\sim (1/n- 1/(n+1))$) is the relevant one for $g_1^{(e.m.)}$. It
turns out \cite{phd,arg} that the coefficient is the same if electroweak
contributions to $g_1$ (e.g.\ $g_1^{\nu p}$) from transitions between
massless ($\lambda=0$) quarks $q\rightarrow q'$ are considered. For the
corrections to the
structure functions $g_3$, $g_4/2x$ one finds\footnote{Eqs.\ (29) are
actually independent of the regularization scheme chosen even in schemes
where $\Delta C_q^1 (1) \neq -3C_F/2$.} \cite{phd,arg}
\begin{eqnarray}
\Delta C_q^3 (n) &=& \Delta C_q^1 (n) + C_F
\left( \frac{1}{n} - \frac{1}{n+1} \right) \nonumber \\
\Delta C_q^4 (n) &=& \Delta C_q^3 (n) + 2 C_F \frac{1}{n+1} \:\:\: .
\end{eqnarray}
One notes the striking similarity to the relations between the quark
corrections to the unpolarized structure functions $F_3$, $F_1$,
$F_2/2x$ which is readily explained by the similarity of the corresponding
hadronic tensors. Equation (29) shows that
the Callan-Gross-type relation $g_4=2xg_3$ mentioned earlier
is broken even for massless quarks beyond the LO. However, unlike its
unpolarized analogue,
$F_L \equiv F_2-2x F_1$, which (in the singlet case) also receives
contributions from {\em gluon}-induced ${\cal O}(\alpha_s)$ corrections,
$g_4-2xg_3=0$ is broken only by quark-induced corrections (eq.\ (29))
even in the singlet case, since corrections from incoming gluons
cancel out for massless produced quarks \cite{wir}.

In the case of the CC transition $s \rightarrow c$ we again have to take into
account the mass of the charm quark which has an influence on the
coefficient functions. For this purpose we have calculated the contribution
of the process $W^{+} s \rightarrow c (g)$ with $m_c \neq 0$ to
$g_1$, $g_3$, $g_4$ in $\overline{\mbox{MS}}$ dimensional regularization,
following the techniques developed in \cite{gott}.
The results of our calculation can be found in the
Appendix. It should be noted that the expressions have a smooth limit
$m_c^2/Q^2 \rightarrow 0$ ($\lambda \rightarrow 1$), in which they
reproduce eqs.\ (A.4),(29). From eqs.\ (A.6-A.9) we
immediately read off the ${\cal O}(\alpha_s)$ corrections to the functions
$f_i (\lambda,n)$ introduced in eqs.\ (11-14). For the first moment,
$n=1$, these functions then read in NLO:
\begin{eqnarray}
f_1 (\lambda,1) &=& \lambda \left( 1 - 3 \frac{\alpha_s}{2\pi}C_F
\frac{(1-\lambda)}{\lambda} \ln (1-\lambda) \right) \nonumber \\
f_3 (\lambda,1) &=& \lambda \left( 1 + \frac{\alpha_s}{2\pi}C_F
\frac{1-\lambda}{\lambda^2} \left[ 1-\frac{\lambda}{2} +
\left( \frac{1}{\lambda}-1 -3 \lambda \right) \ln (1-\lambda) \right] \right)
\nonumber \\
f_4 (\lambda,1) &=& 1 \:\:\: .
\end{eqnarray}
The last result that $f_4 (\lambda,1)$ receives no ${\cal O}(\alpha_s)$
corrections also holds for the corresponding function for the
unpolarized structure function $F_2$ \cite{bij},
where it is in accordance with the Adler sum-rule \cite{adler}.
We emphasize that, similar to the unpolarized case \cite{kramer,ccfr},
our results (A.6-A.9) for the contribution of the
transition $s\rightarrow c$ to the spin-dependent structure functions
would enable a determination of the proton's $\overline{\mbox{MS}}$
polarized strange quark distribution via a detection of charmed final
states in polarized CC DIS.

The (de facto) regularization scheme independence of the relation
$C_q^2 (n) - \Delta C_q^1 (n) =  C_F ( 1/n  + 1/(n+1) )$
immediately implies that the scheme-dependent parts of the
polarized and the unpolarized NS anomalous dimensions
$\Delta \gamma_{NS}^1 (n)$, $\gamma_{NS}^1 (n)$ equal each
other in all schemes. Even more, as was first observed in
\cite{kod} and recently established in more detail in
\cite{willi}, the {\em full} expressions for $\Delta \gamma_{NS}^1 (n)$
and $\gamma_{NS}^1(n)$ are exactly identical. This statement
is correct in all regularization schemes, provided one has taken
care to warrant $\Delta C_q^1 (1)=-3 C_F/2$ in the scheme used,
eliminating, if present, chirality breaking terms as explained above.

There is, however, another subtlety involved in the equality of
$\Delta \gamma_{NS}^1(n)$ and $\gamma_{NS}^1(n)$: As is well-known
[44-46], there is no analytical continuation of the
unpolarized $\gamma_{NS}^1 (n)$ to arbitrary $\:\: n$, needed for
the transformation from Mellin-$n$ space into Bj\o rken-$x$ space,
that reproduces the results for $\gamma_{NS}^1 (n)$ for {\em all}
integer values of $n$. This is not unexpected since the OPE,
first used to derive $\gamma_{NS}^1 (n)$ in $\overline{\mbox{MS}}$
dimensional regularization \cite{flor}, gives only an answer for
{\em even} $n$ if the moments of the NS contribution to the e.m.
structure function $F_2/x$ are considered, hereby artificially
excluding odd values of $n$. Therefore, the analytic
continuation of
$\gamma_{NS}^1 (n)$ only has to correctly reproduce the results for
even values of $n$. On the other hand, as was shown in \cite{nan},
{\em odd} values of $n$ are relevant in the OPE for the NS combination
$F_2^{\bar{\nu} p}/x-F_2^{\nu p}/x$ or for
$F_3^{\bar{\nu} p}+F_3^{\nu p}$, meaning that in this case the
analytic continuation of $\gamma_{NS}^1 (n)$ has to reproduce the results
at these values. These observations fit nicely and consistently to
parton model considerations, where the NS quark combinations $q-q'$ and
$q_v=q-\bar{q}$ can be easily seen to evolve \cite{curci,arep}
with $P_+ \equiv P_{qq}+P_{q\bar{q}}$ and $P_- \equiv P_{qq}-P_{q\bar{q}}$,
respectively, which are different beyond the LO, where $P_{qq}$,
$P_{q\bar{q}}$ are the $q\rightarrow q$ and $\bar{q} \rightarrow q$
NLO NS splitting functions with flavor-non-diagonal contributions
subtracted \cite{curci,arep}. The explicit calculation of
$P_{qq}$, $P_{q\bar{q}}$ \cite{curci} shows that their Mellin-$n$
moments satisfy\footnote{For simplicity we have normalized the $P_{\pm}$
relative to $\gamma_{NS}^1$.}
\begin{equation}
\gamma_{NS}^1 (n) = P_{qq} (n) + (-1)^n P_{q\bar{q}} (n) \:\:\: ,
\end{equation}
which means that the analytic continuation of $\gamma_{NS}^1 (n)$
which reproduces the values of $\gamma_{NS}^1 (n)$ for {\em even}
$n$ equals the combination $P_+ (n)$ for {\em arbitrary} $n$,
whereas the other analytic continuation of $\gamma_{NS}^1 (n)$,
which is correct for {\em odd} $n$, corresponds to $P_- (n)$.
In this way, the parton results of \cite{curci} provide the rule
for the analytic continuation of the OPE results. The essence of all
this is that the moments of the combination $F_2^{ep}/x- F_2^{en}/x$, or,
more in general, the unpolarized $A_3 (n,Q^2)$, $A_8 (n,Q^2)$ (defined
in analogy to eq.\ (7)) evolve
with $P_+ (n)$, whereas, e.g., the moments of $F_2^{\bar{\nu} p}/x-
F_2^{\nu p}/x$, $F_3^{\bar{\nu} p}+F_3^{\nu p}$ (which consist of pure
valence, $u_v (n,Q^2)$, $d_v (n,Q^2)$), evolve with $P_- (n)$.

The important difference in the polarized case is that the relevance
of even and odd $n$ in the OPE and for the analytic continuation is
reversed here.
As was shown in \cite{wray,ahr}, {\em odd} $n$ contribute
in the OPE analysis to the combinations $(g_1^{ep\: (e.m.)}-g_1^{en
\: (e.m.)}) (n,Q^2)$, $(g_1^{\bar{\nu}p}+g_1^{\nu p}) (n,Q^2)$,
$(g_3^{\nu p}-g_3^{\nu n}) (n,Q^2)$, $(g_4^{\nu p} -
g_4^{\nu n}) (n-1,Q^2)$, whereas {\em even} $n$ are relevant, e.g., for
$(g_1^{\bar{\nu}p}-g_1^{\nu p})(n,Q^2)$, $(g_3^{\bar{\nu}p}+
g_3^{\nu p})(n,Q^2)$, $(g_4^{\bar{\nu}p}+g_4^{\nu p})(n-1,Q^2)$.
In terms
of the polarized NS quark distribution combinations this means
that $\Delta A_3 (n,Q^2)$, $\Delta A_8 (n,Q^2)$ (as defined in (7)) evolve
with $P_- (n)$ and the polarized valence densities $\Delta u_v (n,Q^2)$,
$\Delta d_v (n,Q^2)$ with $P_+ (n)$ \cite{grsv}. This situation is summarized
by the relations $\Delta P_{qq}=P_{qq}$, $\Delta P_{q\bar{q}}
=-P_{q\bar{q}}$ for the polarized analogues,
$\Delta P_q$ {\raisebox{-0.6mm}
{$\!\!\! \stackrel{\scriptscriptstyle (-)}{\scriptstyle q} $}},
of $P_q$ {\raisebox{-0.6mm}
{$\!\!\! \stackrel{\scriptscriptstyle (-)}{\scriptstyle q} $}}.
\section{Numerical results}
We are now equipped with all ingredients for a consistent NLO analysis
of the spin-dependent NS structure functions. Let us consider the first
moment of the NS combinations in eqs.\ (10,15-20). To begin with, we
recall that the first moment $P_-(1)=\gamma_{NS}^1 (1)=
\Delta \gamma_{NS}^1 (1)$ vanishes \cite{curci}.
In the unpolarized case this is in accordance with the Adler sum-rule
\cite{adler} and the conservation of the number of
valence quarks. In the polarized case it means that
$\Delta A_3 (n,Q^2)$ and $\Delta A_8 (n,Q^2)$ for $n=1$ correctly do not
evolve with $Q^2$, as required by the conservation of the NS axial
current (see above):
\begin{equation}
\Delta A_{3,8} (1,Q^2) = \Delta A_{3,8}(1,Q_0^2)  \:\:\: .
\end{equation}
In contrast to this, the first moment of $P_+$ is non-zero, which
means that the first moment of the polarized valence densities
evolves with $Q^2$ beyond the LO:
\begin{equation}
\Delta q_v (1,Q^2) = \left( 1+\frac{\alpha_s (Q^2)-\alpha_s (Q_0^2)}{4\pi}
\frac{P_+ (1)}{2\beta_0} \right) \Delta q_v (1,Q_0^2) \:\:\: ,
\end{equation}
with \cite{curci} $P_+ (1) = 4 C_F (C_F-C_A/2)
(-13+12 \zeta (2) - 8 \zeta (3)) \approx 2.5576$, and where we have
used eq.\ (24) and the fact that the first moment of the LO NS anomalous
dimension vanishes, $\gamma_{NS}^0 (1)=0$ \cite{ahr,ap}. This yields
the following sum-rules for the
first moments of the polarized NS structure functions to ${\cal O}
(\alpha_s)$:
\begin{eqnarray}
&&\left( g_1^{ep\: (e.m.)} - g_1^{en \: (e.m.)}\right) (1,Q^2) =
\frac{1}{6} \left( 1-\frac{\alpha_s (Q^2)}{\pi} \right)
\Delta A_3 (1,Q_0^2)  \\
\nonumber
&&g_3^{ep,NC} (1,Q^2) = \frac{1}{4}
\left( 1-\frac{2\alpha_s (Q^2)}{3\pi} +
\frac{\alpha_s (Q^2)-\alpha_s (Q_0^2)}{4\pi} \frac{P_{+}(1)}
{2\beta_0} \right) \left(
\frac{2}{3} \Delta u_v  + \frac{1}{3} \Delta d_v \right) (1,Q_0^2) \\[0.0ex]
\\
&&\frac{1}{2} g_4^{ep,NC} (0,Q^2) = \frac{1}{4}
\left( 1+ \frac{\alpha_s (Q^2)-\alpha_s (Q_0^2)}{4\pi} \frac{P_{+}(1)}
{2\beta_0} \right) \left(
\frac{2}{3} \Delta u_v  + \frac{1}{3} \Delta d_v \right) (1,Q_0^2) \\
&&\left( g_1^{\bar{\nu}p} - g_1^{\nu p} \right) (1,Q^2)
= \left(\! 1-\frac{\alpha_s (Q^2)}{\pi} +
\frac{\alpha_s (Q^2)-\alpha_s (Q_0^2)}{4\pi} \frac{P_{+}(1)}
{2\beta_0}\! \right)\!\left( \Delta u_v  -\Delta d_v \right)\!(1,Q_0^2) \\
&&\left( g_3^{\nu p} - g_3^{\nu n} \right) (1,Q^2)
= \left( 1-\frac{2\alpha_s (Q^2)}{3\pi} \right) \Delta A_3 (1,Q_0^2) \\
&&\frac{1}{2} \left( g_4^{\nu p} - g_4^{\nu n} \right) (0,Q^2)
= \Delta A_3 (1,Q_0^2) \\
\nonumber
&&\left( g_3^{\bar{\nu}p} + g_3^{\nu p} \right) (1,Q^2)
= -\left( 1-\frac{2\alpha_s (Q^2)}{3\pi} +
\frac{\alpha_s (Q^2)-\alpha_s (Q_0^2)}{4\pi} \frac{P_{+}(1)}
{2\beta_0} \right) \left( \Delta u_v + \Delta d_v \right) (1,Q_0^2) \\[0.0ex]
\\
&&\frac{1}{2} \left( g_4^{\bar{\nu}p} + g_4^{\nu p} \right) (0,Q^2)
= - \left( 1+\frac{\alpha_s (Q^2)-\alpha_s (Q_0^2)}{4\pi} \frac{P_{+}(1)}
{2\beta_0} \right) \left( \Delta u_v + \Delta d_v \right) (1,Q_0^2) \\
&&\left[ 9 \left( g_1^{ep \: (e.m.)} + g_1^{en (e.m.)}  \right) -
\frac{6}{2+f_1 (\lambda,1)}
\left( g_1^{\nu p} + g_1^{\nu n}  \right) \right] (1,Q^2)  \nonumber \\
&&\hspace*{6cm} = \frac{1}{2}\;\frac{5f_1(\lambda,1)-2}{f_1(\lambda,1)+2}
\left( 1-\frac{\alpha_s (Q^2)}{\pi} \right)
\Delta A_8 (1,Q_0^2)\;\;\;,
\end{eqnarray}
etc. It should be noted that eq.\ (20) receives singlet contributions
beyond the leading order, therefore we have not written down
this equation any more.
Eqs.\ (34-42) show how in principle measurements of the first moments of
polarized NS structure functions even at large $Q^2$ can serve to
independently determine the combinations $(\Delta u_v \pm
\Delta d_v ) (1,Q_0^2)$, $\Delta A_3 (1,Q_0^2)$ and $\Delta A_8 (1,Q_0^2)$.
This is particularly interesting considering the question raised earlier
of which combination of polarized parton distributions can be related
to the $F$, $D$ values measured in baryonic $\beta$-decays.
To simplify the discussion, we follow the recent NLO analysis \cite{grsv}
to assume that at the low input scale
$Q_0^2=0.34$ GeV$^2$ $(\equiv \mu_{NLO}^2$ \cite{grsv})
we can neglect any effects of $SU(2)$ isospin
breaking in relating $\beta$-decay matrix elements of charged and
neutral currents as well as $SU(2)_f$ breaking in the proton's
polarized sea. We then have $\Delta A_3 (1,Q_0^2)=(\Delta u_v -
\Delta d_v) (1,Q_0^2)= F+D$, and the rhs of eqs.\ (34,37-39)
are completely specified, leading to unique predictions for the
combinations of
structure functions on the lhs in NLO of QCD. The first of these
is of course the well-known Bj\o rken sum-rule \cite{bj} to
${\cal O}(\alpha_s)$ \cite{kod}\footnote{Note that
actually the corrections to ${\cal O}(\alpha_s^2)$, ${\cal O}(\alpha_s^3)$
to this sum-rule are known \cite{lv}.}. The results for eqs.\ (34,37-39)
are displayed in Fig.\ 1 as functions of $Q^2$, where we have used
\cite{grsv} $\Lambda_{\overline{\rm{MS}}}^{(f=4)}=200$ MeV.
To account for $SU(3)$
breaking effects we parametrize the input quantities appearing
on the rhs of eqs.\ (40-42) in the following way:
\begin{eqnarray}
\Delta A_8 (1,Q_0^2)&=& (3 F-D) (1-\epsilon_1) \\
\left( \Delta  u_v  + \Delta d_v  \right)
(1,Q_0^2) &=& \frac{1}{1-\epsilon_2} (3F-D)  \:\:\: ,
\end{eqnarray}
which yields $(3F-D)/2(1-\epsilon_2)+(F+D)/6$ for the
combination $(2\Delta u_v/3+\Delta d_v/3)(1,Q_0^2)$ in eqs.\ (35,36).
Eqs.\ (43,44) are general enough to take into account all
possible sources of $SU(3)$ breaking: $\epsilon_1$ determines the
deviation of the first moment $\Delta A_8 (1,Q_0^2)$ from the
value $3F-D$ obtained from hyperon $\beta$-decays. Such a deviation
will occur if the use of $SU(3)$ symmetry for relating the matrix
elements of charged and neutral axial currents is not justified.
In this case, $\epsilon_1$ could be significantly different from
zero, even such that only the {\em valence} quarks contribute
to $3F-D$ \cite{lip}. This possibility is taken into account
by the parameter $\epsilon_2$ which would vanish in the latter case. From
the definition of $\Delta A_8$ one furthermore sees that
$\epsilon_1$ and $\epsilon_2$ together determine the amount of
$SU(3)_f$ breaking in the proton's polarized sea:
\begin{equation}
2 \frac{\Delta \bar{u}+\Delta \bar{d}-2 \Delta \bar{s}}
{\Delta u_v+\Delta d_v} (1,Q_0^2)
= (1-\epsilon_1)(1-\epsilon_2) -1 \:\:\: .
\end{equation}
Fig.\ 1 shows our predictions for the NS structure functions of
eqs.\ (35,36,40-42) for the conceivable choices \cite{grsv} $\epsilon_1 =0$,
$\epsilon_2 =0.105$ and $\epsilon_1 =0.40$, $\epsilon_2=0$.
It becomes obvious that the effects of changes in the $\epsilon_i$ are
larger than the present experimental $4\%$-uncertainty
\cite{close} in the value for $3F-D$ (see eq.\ (4)) and that therefore
a measurement of the quantities shown would help to decide about
the amount of $SU(3)$ breaking. In particular, the parameter $\epsilon_2$
could be determined in NC,CC experiments with polarized beams at
HERA \cite{barber} via a measurement of $g_4^{ep,NC}(0,Q^2)$ or
$(g_4^{\bar{\nu}p}+g_4^{\nu p})(0,Q^2) \equiv
(g_4^{e^{-}p,CC}+g_4^{e^{+}p,CC})(0,Q^2)$ (or their $g_3$-analogues).
Using the full Mellin-$n$-dependent expression for $\Delta C_q^4 (n)$
from eqs.\ (29),(A.4) in eq.\ (22), we can obtain NLO predictions
for the Bj\o rken-$x$ dependence of the latter structure functions:
\begin{equation}
\left.
\begin{array}{l}
g_4^{ep,NC}(n-1,Q^2) \\
(g_4^{e^{-}p,CC}+g_4^{e^{+}p,CC})(n-1,Q^2)
\end{array}\!\!
\right\}\! =\!
\left( 1+ \frac{\alpha_s(Q^2)}{2\pi} \Delta C_q^4 (n) \right)
\!\left\{
\begin{array}{l}
\frac{1}{2} \left( \frac{2}{3} \Delta u_v + \frac{1}{3}
\Delta d_v \right) (n,Q^2) \\
(-2) \left( \Delta u_v + \Delta d_v \right) (n,Q^2)\;\;,
\end{array}
\right.
\end{equation}
where, again, the polarized valence quark densities are to be evolved
according to eq. (24) with the correct analytic continuation
$P_+ (n)$ of the $\gamma_{NS}^1 (n)$ found in \cite{curci,flor,flor1}.
The results for $g_4^{ep,NC}(x,Q^2)$ and
$(g_4^{e^{-}p,CC}+g_4^{e^{+}p,CC})(x,Q^2)$ at $Q^2=1000$ GeV$^2$,
found after Mellin-inverting eq. (46), are shown in Fig.\ 2,
where for the polarized input valence densities $\Delta q_v (x,Q_0^2)$
at $Q_0^2 =0.34$ GeV$^2$ we  have used the two sets determined
in ref. \cite{grsv}. Both sets give a very good description of all
existing data on deep-inelastic spin asymmetries in the valence
region
$x\stackrel{>}{\scriptscriptstyle \sim}0.2$, but they differ
in the assumptions made about the role of $SU(3)_f$ symmetry breaking
effects and therefore have different first moments
\cite{grsv}, corresponding to the $\epsilon_1$, $\epsilon_2$ values used
in Fig.\ 1.
Thus the variation in the results shown in Fig.\ 2 for the different
sets of polarized valence input densities reflects the present theoretical
uncertainty in the predictions. Conversely, Fig.\ 2 shows that also a
measurement of $g_4^{ep,NC}(x,Q^2)$ and $(g_4^{e^{-}p,CC}+
g_4^{e^{+}p,CC})(x,Q^2)$ for $x \leq 0.2$ at HERA could help to shed
light on the importance of $SU(3)_f$ symmetry breaking.

The different evolution of the polarized valence quark densities
and the combination $\Delta A_3$ beyond the LO induces
a dynamical breaking of the $SU(2)_f$ symmetry in the
proton's polarized sea\footnote{There is also a dynamical breaking
of $SU(3)_f$ symmetry in the sea induced by $\Delta A_8$. We do not
pursue this effect since it is most probably dominated by the $SU(3)_f$
breaking in the nonperturbative input (see eq.\ (41))
due to the larger strange mass.}. Eq.\ (24) and our considerations
concerning the analytic continuation of $\gamma_{NS}^1 (n)$
predict
\begin{eqnarray}
\nonumber
2 \left( \Delta \bar{u}-\Delta \bar{d} \right) (n,Q^2)
&=& \frac{\left(\alpha_s (Q^2)-\alpha_s (Q_0^2)\right)}{4 \pi}
\;\frac{\left(P_- (n)-P_+ (n)\right)}{2 \beta_0}\\
&&\times
\left( \frac{\alpha_s (Q^2)}{\alpha_s (Q_0^2)}
\right)^{\gamma_{NS}^0(n)/2\beta_0} \left( \Delta u_v -
\Delta d_v \right) (n,Q_0^2)  \:\:\: ,
\end{eqnarray}
where we have again assumed the nonperturbative input at $Q_0^2$
to be $SU(2)_f$-symmetric, $\Delta \bar{u}(n,Q_0^2) =
\Delta \bar{d} (n,Q_0^2)$. Using again the
polarized input valence distributions of ref.\ \cite{grsv} at $Q_0^2 =0.34$
GeV$^2$, we obtain a prediction for $(\Delta \bar{u}-\Delta \bar{d})
(x,Q^2=10 \mbox{GeV}^2)$ which is shown in Fig.\ 3. Both sets of
polarized input valence densities considered in \cite{grsv} lead to
entirely indistinguishable results, since only the input
combination $(\Delta u_v -\Delta d_v)(x,Q_0^2)$ is needed here whose first
moment is fixed by the value $F+D$ (eq. (3)) in both cases.
It should be noted that such a dynamical breaking of $SU(2)_f$ symmetry
in the sea
induced by two-loop evolution was considered in the unpolarized case
in ref.\ \cite{rs}, where it was found to be very small.
The results in the polarized case differ in sign
(due to the interchange $P_+ \leftrightarrow P_-$) and slightly
in magnitude (due to the polarized valence input instead of the
unpolarized one) from the unpolarized results. However, the {\em relative}
effect of $SU(2)_f$ breaking is much larger in the polarized case
since the polarized sea densities are probably much smaller than
their unpolarized counterparts. Even so, when taking the first moment,
one obtains\footnote{This number depends quite crucially on the
value chosen for the input scale $Q_0$. Taking, e.g., $Q_0^2=1$
GeV$^2$ the result in eq. (48) is reduced by a factor 3.}
\begin{eqnarray}
2 \left( \Delta \bar{u}-\Delta \bar{d} \right) (1,Q^2=10 \mbox{GeV}^2)
&=& - \frac{\left(\alpha_s (Q^2)-\alpha_s (Q_0^2)\right)}{4 \pi}\;
\frac{P_+ (1)}{2 \beta_0} (F+D) \nonumber \\
&\approx & 0.006 \:\:\: ,
\end{eqnarray}
which means that unless the input at $Q_0^2$ strongly breaks
$SU(2)_f$ symmetry the effect of the breaking is probably small
compared to the size of $|\Delta \bar{u}(1,Q^2)|$, $|\Delta \bar{d}(1,Q^2)|$
which might well be of the order
$\stackrel{>}{\scriptscriptstyle{\sim}}0.05$
\cite{grsv}. It is straightforward to introduce parameters
$\delta_1$, $\delta_2$ in analogy to $\epsilon_1$, $\epsilon_2$,
which would parametrize genuine $SU(2)$ breaking effects in the first
moment of the polarized sea and in the relation between charged and
neutral axial current $\beta$-decay matrix elements. Measurements of the
first moment of the structure functions in eqs.\ (34,37-39) (see also
Fig.\ 1) would then allow to determine these parameters and to
pin down $SU(2)$ breaking effects.
\section{Summary and conclusions}
We have performed a detailed study of spin-dependent non-singlet
structure functions
in the framework of a complete and consistent NLO QCD calculation.
Our analysis is based on a careful discussion of the calculation of the
${\cal O}(\alpha_s)$ corrections to the structure functions, in which
we have examined the regularization scheme dependence of the NS coefficient
function $\Delta C_q^1$ for $g_1$ with respect to the constraints
imposed by axial current conservation. We have also shown how to
correctly take into account the two-loop evolution of polarized
NS quark combinations. A further ingredient of our study is the full
inclusion of the charm mass effects in the charged current
$s\rightarrow c$ contributions to polarized electroweak structure
functions.

Our numerical analysis has revealed that conceivable measurements of
spin-dependent NS structure functions at HERA or in
$\stackrel{\scriptscriptstyle (-)}{\nu}\!$ scattering
experiments off polarized nucleon targets
would serve to improve our understanding of the relations between
the first moment of NS combinations of polarized quark densities
and the $F$, $D$ values extracted from hyperon-$\beta$-decays,
and would also shed light on $SU(2)_f$, $SU(3)_f$ breaking in the nucleon's
polarized sea. Finally, we have also shown that the latter symmetries
are dynamically broken by NLO evolution in the NS sector.
\section*{Acknowledgements}
This work has been supported in part by the
'Bundesministerium f\"{u}r Bildung, Wissenschaft, Forschung und
Technologie' (former BMFT), Bonn.
%
\section*{Appendix}
\setcounter{equation}{0}
\renewcommand{\theequation}{A.\arabic{equation}}
In this Appendix we list the results for the polarized coefficient
function $\Delta C_q^1 (n)$ using various regulators in its calculation
from the process $\vec{\gamma}^{\ast}\vec{q} \rightarrow q (g)$.
In all cases we have chosen to just subtract the collinear
pole contribution, which
is then factorized into the (bare) quark distributions. The
singular terms are of the forms $\gamma_{NS}^0 (n) \ln (Q^2/|m^2|)$
(if some mass or off-shellness $\sqrt{|m^2|}$ is used as the regulator) or
$\gamma_{NS}^0 (n) (-1/\hat{\epsilon})$ (in dimensional regularization
in the $\overline{\mbox {\rm MS}}$-scheme). Since the first moment of
$\gamma_{NS}^0 (n)$ vanishes \cite{ahr,ap}, the pole contribution drops out
from the more important first moment anyway. The results in Mellin-$n$ space
below can be easily transformed into Bj\o rken-$x$ space
with the help of the Appendix of ref.\ \cite{aem2}.
\begin{description}
\item[Off-shell massless quarks, on-shell gluons:] This calculation
corresponds to the one of ref.\ \cite{kod}, but our results slightly differ
by a term $C_F ( -1+2/(n+1))$ (which vanishes for $n=1$) due to
the specific operator normalization chosen in \cite{kod} (see the
Appendix A.2 of ref.\ \cite{kod} for details). For the result in the
unpolarized case ($F_2$) see \cite{aem1}.
\begin{equation}
\Delta C_q^1 (n) = C_F \left[ -\frac{3}{2n}+\frac{2}{n+1}+
\frac{2}{n^2}-\frac{2}{(n+1)^2}+\frac{3}{2} \sum_{j=1}^n \frac{1}{j}-
4\sum_{j=1}^n \frac{1}{j^2} \right]     \:\:\: .
\end{equation}
\item[On-shell massless quarks, off-shell gluons:] The
$\Delta C_q^1 (n)$ for this calculation can be obtained from \cite{guil}.
For the unpolarized case see \cite{kubar}.
\begin{eqnarray}
\Delta C_q^1 (n) &=& C_F \left[ -\frac{9}{4}-\frac{3}{2n}+\frac{3}{n+1}+
\frac{2}{n^2}-\frac{1}{(n+1)^2} \right. \nonumber \\
&&\left. +\left( \frac{3}{2} -\frac{1}{n(n+1)} \right)
\sum_{j=1}^n \frac{1}{j}-
4\sum_{j=1}^n \frac{1}{j^2} +2 \sum_{j=1}^n\frac{1}{j}
\sum_{k=1}^j \frac{1}{k} \right]     \:\:\: .
\end{eqnarray}
\item[On-shell massive quarks, off-shell gluons:] In this regularization
we obtain:
\begin{eqnarray}
\Delta C_q^1 (n) &=& C_F \left[ -\frac{5}{2}-\frac{5}{2n}+\frac{2}{n+1}+
\frac{1}{n^2}-\frac{2}{(n+1)^2} \right. \nonumber \\
&&\left. +\left( \frac{7}{2} +\frac{1}{n(n+1)} \right) \sum_{j=1}^n
\frac{1}{j}- 2\sum_{j=1}^n \frac{1}{j^2} - 2 \sum_{j=1}^n\frac{1}{j}
\sum_{k=1}^j \frac{1}{k} \right]     \:\:\: .
\end{eqnarray}
Note that $\Delta C_q^1 (1) = -7 C_F/2$ in this scheme.  See refs.\
\cite{calvo,kod} for the corresponding unpolarized result.
\item[Dimensional regularization:]
Using the $\gamma_5$-prescription of \cite{cfh} (or its
more systematic and consistent generalization \cite{korn})
one obtains in the $\overline{\mbox{MS}}$-scheme \cite{rat,willi}:
\begin{eqnarray}
\Delta C_q^1 (n) &=& C_F \left[ -\frac{9}{2}+\frac{1}{2n}+\frac{1}{n+1}+
\frac{1}{n^2} \right. \nonumber \\
&&\left. +\left( \frac{3}{2} -\frac{1}{n(n+1)} \right) \sum_{j=1}^n
\frac{1}{j}- 2\sum_{j=1}^n \frac{1}{j^2} + 2 \sum_{j=1}^n\frac{1}{j}
\sum_{k=1}^j \frac{1}{k} \right]     \:\:\: .
\end{eqnarray}
The same result in dimensional regularization was found earlier in
\cite{kod2} without specifying the $\gamma_5$-prescription.
However, using the original scheme of 't Hooft and Veltman \cite{tv}
and Breitenlohner and Maison \cite{bm} (or the equivalent one of
refs.\ \cite{del,lv}), one finds \cite{alex,phd} an additional term
\begin{equation}
\Delta C_q^1 (n)^{\rm{HVBM}} = \Delta C_q^1 (n)^{\rm{(A.4)}} -
4 C_F \left( \frac{1}{n} - \frac{1}{n+1} \right) \:\:\: ,
\end{equation}
which leads to $\Delta C_q^1 (1)=-7 C_F/2$. For the unpolarized case
see \cite{aem2,bb}.
\end{description}

We finally present our results for the coefficient functions
$\Delta \tilde{C}_q^i$ for
$g_1$, $g_3$, $g_4/2x$ for the transition $s\rightarrow c$, fully taking into
account the effects due to the charm quark mass. The calculation was
performed in $\overline{\mbox{MS}}$ dimensional regularization in the
$\gamma_5$-scheme of ref.\ \cite{korn}, choosing the axial vertex as
the reading point. Our Bj\o rken-$x$ space results for
$\Delta \tilde{C}_q^1$, $\Delta \tilde{C}_q^3$, $\Delta \tilde{C}_q^4$
fully agree with those of ref.\ \cite{gott} for the unpolarized $h_{3,q}$
$h_{1,q}$, $h_{2,q}$ (for $F_3$, $F_1$, $F_2/2x$), respectively,
after eliminating an error in the coefficient $A_2$ in that paper
which should read $K_A$ instead of $K_A/2$. The differences
$\Delta \tilde{C}_q^4 - \Delta \tilde{C}_q^1$,
$\Delta \tilde{C}_q^4 - \Delta \tilde{C}_q^3$ which are regularization
scheme independent, are in agreement with the results of \cite{bij}
for the corresponding differences in the unpolarized case. We note
that the results of ref.\ \cite{kramer} seem in slight disagreement
with both \cite{gott} (even after correction of the above mentioned
error) and \cite{bij} and also with our calculation in this respect.
Here we present
the Mellin-$n$ moments of our results. For this purpose
it is convenient to present the moments for the differences
$\Delta \tilde{C}_q^i (n,\lambda)-\Delta C_q^i (n)$, where the
$\Delta C_q^i(n)$ are the (usual) massless coefficient functions
given in eqs.\ (29), (A.4), and $\lambda=Q^2/(Q^2+m_c^2)$.
Defining the sum
\begin{displaymath}
{\cal S}_{\alpha} (n,\lambda) \equiv \lambda^{-\alpha n}
\left[ \ln (1-\lambda) + \sum_{j=1}^n \frac{\lambda^{\alpha j}}{j} \right]
\:\:\: ,
\end{displaymath}
and \cite{gott}
\begin{displaymath}
K_A (\lambda) \equiv \frac{1-\lambda}{\lambda} \ln (1-\lambda) \:\:\: ,
\end{displaymath}
we find:
\begin{eqnarray}
&&\!\!\!\!\!\!\!\!\Delta \tilde{C}_q^1 (n,\lambda)-\Delta C_q^1 (n) =
C_F \left[
-K_A (\lambda)+2 \sum_{i=1}^n \frac{1}{i} \left( {\cal S}_0 (i,\lambda)
-{\cal S}_1 (i,\lambda) \right) \right. \nonumber \\
&&\!\!\!\!\!\!\!\!
\hspace*{3cm} + \frac{(n-1)(n+2)}{2n(n+1)}
\left( {\cal S}_0 (n,\lambda) -
{\cal S}_1 (n,\lambda) \right) - \frac{n(n-1)}{2(n+1)}
\frac{1-\lambda}{\lambda} {\cal S}_1 (n,\lambda) \nonumber \\
&&\!\!\!\!\!\!\!\!
\hspace*{3cm} - \left. \left( \frac{3}{2} + \frac{1}{n+1} -2 \sum_{j=1}^n
\frac{1}{j} \right) \ln \lambda \right]     \\
&&\!\!\!\!\!\!\!\!
\left( \Delta \tilde{C}_q^3 - \Delta \tilde{C}_q^1 \right)(n,\lambda) -
\left( \Delta C_q^3 - \Delta C_q^1  \right) (n) = C_F \Bigg[
\frac{1-\lambda}{\lambda}\frac{1}{n+1}+
\frac{(1-\lambda)^2}{\lambda^2} {\cal S}_1 (n,\lambda) \Bigg]\\
&&\!\!\!\!\!\!\!\!
\left( \Delta \tilde{C}_q^4 - \Delta \tilde{C}_q^3 \right)(n,\lambda) -
\left( \Delta C_q^4 - \Delta C_q^3 \right) (n) = C_F \Bigg[
K_A (\lambda) - \frac{(1-\lambda^2)(1-2 \lambda)}{\lambda^2}
{\cal S}_1 (n,\lambda) \nonumber \\
&&\!\!\!\!\!\!\!\!
\hspace*{3cm} - (1-\lambda) \left( \frac{2}{n} + \frac{1}{\lambda (n+1)}
\right) \Bigg] \:\:\: .
\end{eqnarray}
The last term in eq.\ (A.6) which contains the LO $\gamma_{NS}^0 (n)$
\cite{ahr,ap} is introduced if one chooses the scale $Q^2$
as the factorization
scale \cite{gott}.
It should be noted that, like in the LO (see eqs.\ (11-14)),
an additional factor of $\lambda^n$
($\lambda^{n-1}$) is needed to calculate the contribution to
the structure functions $g_1 (n,Q^2)$, $g_3 (n,Q^2)$
($g_4 (n-1,Q^2)/2$).
\newpage
\newpage
\section*{Figure Captions}
\begin{description}
\item[Fig. 1] Predictions for the $Q^2$-evolution of the first moments of
the various NS combinations of polarized structure functions as given
in eqs.\ (34)-(42) for two conceivable choices of $SU(3)_f$ breaking
parameters $\epsilon_1$, $\epsilon_2$ in eqs.\ (43),(44).
The input scale for the evolution, $Q_0^2=0.34\;{\rm{GeV}}^2$, was chosen
according to ref.\ \cite{grsv}, and $\alpha_s(Q^2)$ was calculated from
eq.\ (23) with $\Lambda_{\overline{\rm{MS}}}$ from \cite{grsv}.
\item[Fig. 2] Predictions for the NC and CC non-singlet
structure functions
(cf.\ eq.\ (46)) $g_4^{ep,NC}(x,Q^2)$ and $g_4^{ep,CC}(x,Q^2) \equiv
\left( g_4^{e^-p,CC}+g_4^{e^+p,CC} \right)(x,Q^2)$,
respectively, as measurable in a future
polarized $e^-p/e^+p$ collider mode of HERA \cite{barber}.
For the predictions we have used the two sets of polarized input
valence densities suggested in \cite{grsv} which correspond to the
$SU(3)_f$ breaking parameters introduced in Fig.\ 1.
\item[Fig. 3] Prediction for the dynamical $SU(2)_f$ breaking of the
proton's polarized sea \linebreak
$\left(\Delta\bar{u}-\Delta\bar{d}\right)(x,Q^2)$
at $Q^2=10\;{\rm{GeV}}^2$ according to eq.\ (47). The nonperturbative
valence input $\left(\Delta u_v -\Delta d_v \right)(x,Q_0^2)$
at $Q^2_0=0.34\;{\rm{GeV}}^2$
was taken from
the analysis in ref.\ \cite{grsv}. For comparison the dashed line shows the
averaged sea density $-\Delta\bar{q}(x,Q^2)\equiv
-\left(\Delta\bar{u}(x,Q^2)+\Delta\bar{d}(x,Q^2)\right)/2$ determined
within the 'standard scenario' of ref.\ \cite{grsv}.
\end{description}
\newpage
\pagestyle{empty}

\vspace*{-1cm}
\hspace*{-0.7cm}
\epsfig{file=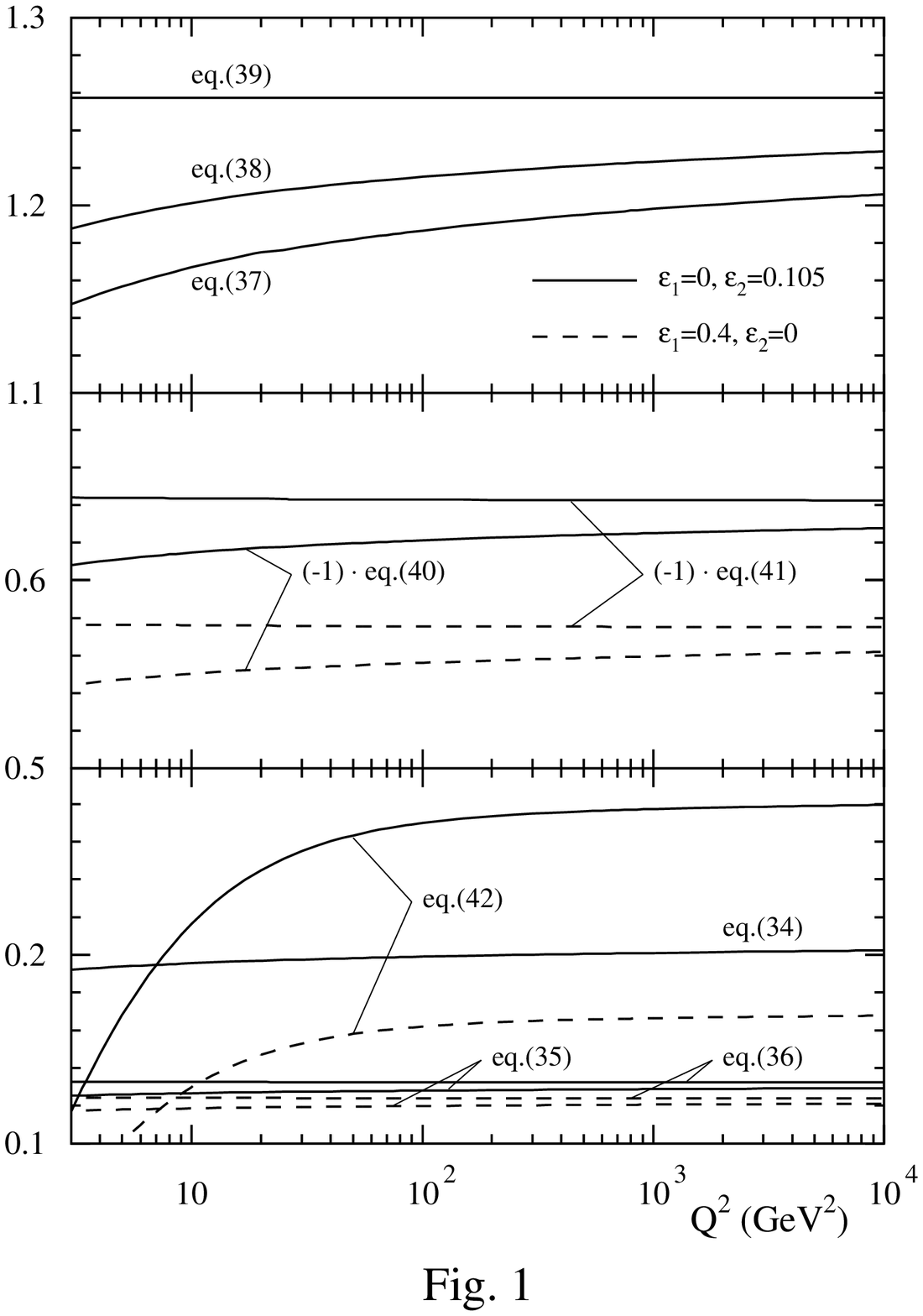}

\newpage
\vspace*{1cm}
\hspace*{-1.4cm}
\epsfig{file=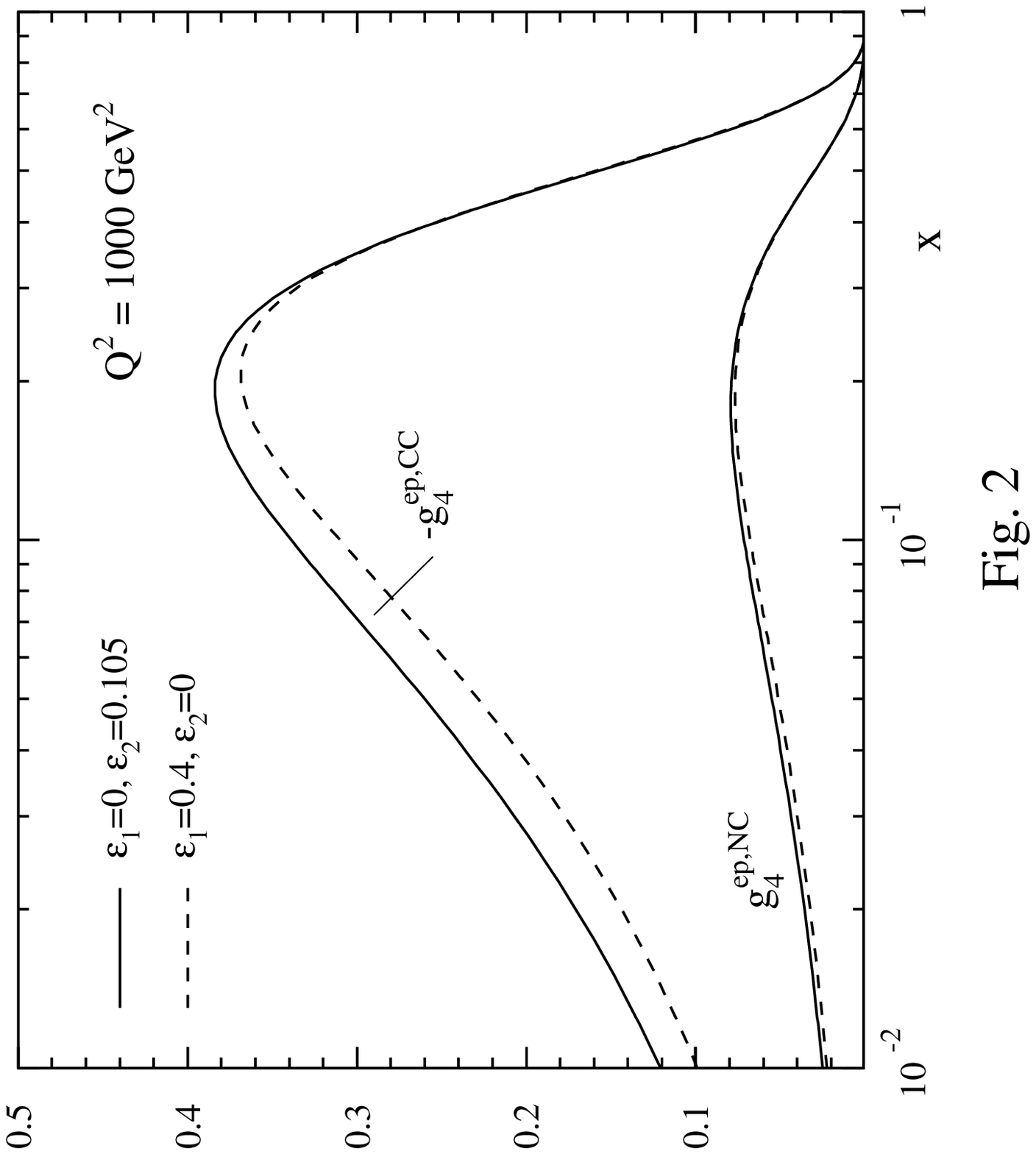,angle=90}

\newpage
\vspace*{1cm}
\hspace*{-1.4cm}
\epsfig{file=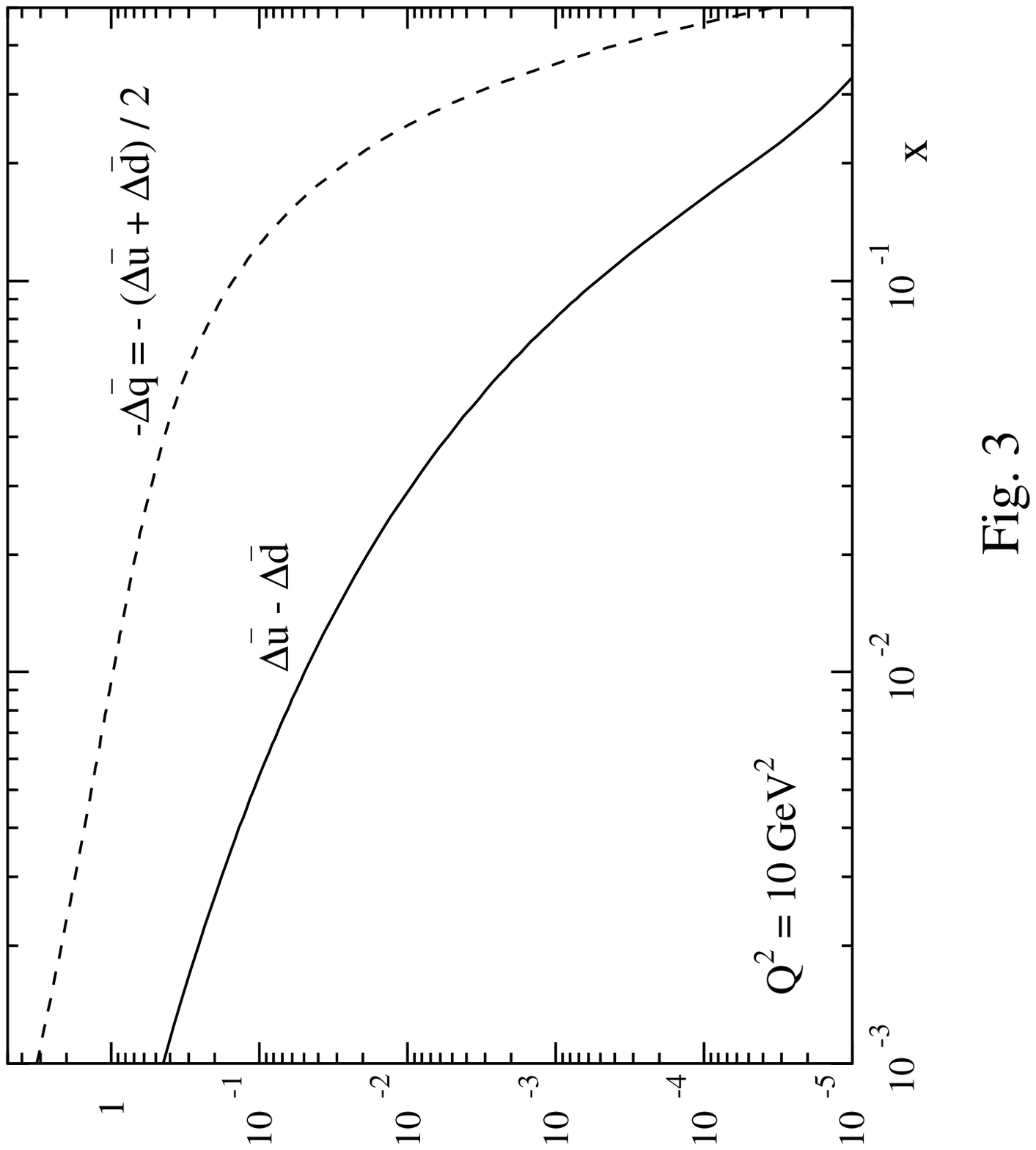,angle=90}

\end{document}